# Putting the dynamic pathosome in practice: a novel way of analyzing longitudinal data

## Authors

Peter Lenart[1,2*], Daniela Kuruczova[1,3*], Lubomír Kukla[1], Martin Scheringer[1,4], Julie Bienertova-Vasku[1]

[1] *Research Centre for Toxic Compounds in the Environment, Faculty of Science, Masaryk University, Kamenice 5, building A29, 625 00, Brno, Czech Republic*

[2] *Department of Experimental Biology, Faculty of Science, Masaryk University, Kamenice 753/5, CZ-625 00 Brno, Czech Republic*

[3] *Department of Mathematics and Statistics, Faculty of Science, Masaryk University, Kotlářská 2, 61137, Brno, Czech Republic*

[4] *Institute of Biogeochemistry and Pollutant Dynamics, ETH Zurich, Universitätstr. 16, CH 8092, Zürich, Switzerland.*

*Authors contributed equally

## Abstract

Previously we have developed the concept of the dynamic pathosome, which suggests that individual patterns of phenotype development, i.e., phenotypic trajectories, contain more information than is commonly appreciated and that a phenotype's past trajectory predicts its future development. In this article, we present a pathosome-inspired approach to analyzing longitudinal data by functional linear models. We demonstrate how to use this approach and compare it with classical linear models on data from the Czech section of the European Longitudinal Study of Pregnancy and Childhood (ELSPAC). Our results show that functional linear models explain more observed variance in age at menarche from height and weight data than the commonly used approaches. Furthermore, we demonstrate that functional linear models can be used to identify crucial time points that can be used to create linear models achieving almost the same performance as functional linear models. In addition, we use data from the Berkeley growth study (BGS) to demonstrate that growth trajectories from birth to 15 years can be used to explain 97% of observed variance of height at 18 years, thus supporting the notion that a phenotype's past trajectory affects its future course. Overall, this article presents experimental support for the concept of the dynamic pathosome and presents a method that can be used as a powerful tool for analyzing quantitative longitudinal data.

## Introduction

In our previous work, we have introduced a novel concept of the dynamic pathosome[1,2] and argued that it could explain the development of various phenotypes, including diseases, better than the contemporary framework describing genotype–environment interactions[3,4]. A critical aspect of the dynamic pathosome is that it presumes that individual phenotypes and their combinations are not static but develop over time in a logical and coherent fashion that may be studied and elucidated. For example, before a clinically healthy phenotype change into a disease phenotype, a given individual may develop several transient phenotypes whose trajectory points towards the disease much sooner than a simple cross-sectional look at the symptoms. In general terms, the dynamic pathosome suggests that phenotype trajectories predict future phenotypes. Furthermore, such phenotypic trajectories reflect the history of genotype-environment interactions of a given individual and, thus, contain much more information than is commonly appreciated[1,2].

The extensive requirements for data collection under the framework of the dynamic pathosome complicate its validation. However, two core suggestions derived from the dynamic pathosome could be tested on existing longitudinal datasets provided an appropriate statistical method existed. The first suggestion is that analyzing phenotypic trajectories in longitudinal studies should be superior to standard epidemiological approaches. The second suggestion testable from available data is that the past phenotype trajectory predicts the future development of that particular phenotype. Unfortunately, commonly used statistical methods are no well-suited for the analysis of phenotype trajectories. Therefore, in this work, we present a novel, pathosome-inspired approach to analyzing longitudinal data based on functional linear models. We demonstrate how this approach can be used in practice and compare it against classical approaches used in epidemiology with respect to its ability to predict the age at menarche from data about height and weight. Thereby, we want to test if analyzing phenotypic trajectories confers any notable benefits. Furthermore, we further use this approach to show how it can be employed to predict future phenotype development from past phenotype trajectories.

## Methods

**Study subjects**

**ELSPAC**

Most of the data used in this study were obtained from the Czech branch of the European Longitudinal Study of Pregnancy and Childhood (ELSPAC), initiated by the World Health Organization Regional Office for Europe.[5] All mothers in the Czech section of the study were recruited in the Brno and Znojmo regions of present-day Czechia and were expected to deliver between 1 March 1991 and 30 June 1992. The cohort included 5151 families who completed the initial questionnaire during pregnancy and at least one questionnaire afterwards[6].

Age at menarche was determined from questionnaires completed at ages 11, 13, 15, 18 and 19 during the course of general practitioner appointments and confirmed by questionnaires filled in by subjects' mothers at ages 15, 18 and 19. Questionnaires at ages 11, 13, and 15 contained questions about the precise age at menarche and questionnaires at ages 18 and 19 only asked whether menarche had already occurred. Therefore, if menarche was listed as not having occurred at age 15 but as having occurred already at 18, we excluded these subjects from the analysis.

Height and weight were measured by the subjects' pediatricians at 3, 5, 7, 11, 15, and 18 years.

The age at menarche of all mothers was determined from a questionnaire administered in the prenatal period.

For age at menarche models, only females with complete height and menarche data were selected, 364 subjects in total.

For height at 18-yr models, all subjects with complete height data were selected, 1155 subjects in total.

Ethical approval for the study was obtained from the ELSPAC Ethics Committee and local research ethics committees. Written informed consent was signed by all study participants.

**Berkeley Growth Study**

As a supplementary dataset, we also used freely available growth data from the Berkeley growth study (BGS)[7–9]. The study measured growth data in 93 participants (39 boys and 54 girls) from one year of age until their 18th birthday. Up to two years of age, the height data were collected in quarter-year intervals; then, one year intervals were used up to eight years. From there on, the height was measured in half-year intervals.

**Statistics**

**Standard approach: Linear models**

To compare the standard and functional approach, we fitted several linear models. The calculation was carried out with the statistical software R (version 4.0.3) and its built-in functions for linear models[10].

For one of the fitted models, the time in which girls experienced their fastest upward growth (the peak height velocity) was calculated by fitting the Preece and Baines Model 1[11] using the

Levenberg-Marquardt Nonlinear Least-Squares Algorithm provided by the R library minipack.lm[12].

**Pathosome-inspired approach: functional linear models**

The idea behind the functional linear model is the following: instead of choosing one or a few timepoints, we use the entire growth curve $x(t)$ i.e., continuous information about the child's height throughout time $t$.

A classical multivariate linear model with $k$ selected timepoints is usually expressed as follows:

$$y = \beta_0 + \sum_{j=1}^{k} \beta_j x(t_j) = \beta_0 + \beta_1 x(t_1) + \cdots + \beta_k x(t_k)$$

A significant limitation of this model is that the number of selected timepoints $k$ must be smaller than the number of measurements (i.e., the number of subjects), ideally by a large margin. The second limitation is the multicollinearity, which occurs when the $x(t_j)$ values are highly correlated. Both limitations can be addressed by choosing a low number of suitable timepoints from the growth curve. Naturally, some information contained in the growth curve is lost due to not using all timepoints.

The functional linear model is the natural expansion of the classic linear model and uses the entire growth curve:

$$y = \beta_0 + \int \beta(t) x(t) \, dt$$

Instead of a sum of selected time points and corresponding coefficients, an integral over time is used. The multivariate linear model's limitations are addressed by using basis expansion and roughness penalty[7] and will not be discussed in detail here. The multivariate linear model results are the values of the $k$ coefficients $\beta_j$, which are usually tested against the zero-value null hypothesis. In the linear functional model, the resulting β coefficient is not a number but rather a curve. Instead of hypothesis testing, confidence intervals can be constructed to determine when the function *x(t)* is associated with the outcome variable *y*. Examples and the interpretation of such results will be illustrated further down.

The child's growth is a continuous process, but it is impossible to measure it continuously in practice. Longitudinal data usually record the child's height at pre-specified discrete time points. Several options are available for the reconstruction of continuous growth curves from discrete measurements. One is to fit a parametric model, such as Preece-Baines. The other option is to reconstruct the curves in a non-parametric fashion (i.e., with no prior assumptions about their shape) using kernel or spline smoothing.

The argument about continuity also applies to the weight, but several differences need to be pointed out. While genetic influence is still present, weight is significantly more influenced by external factors. Therefore, only a non-parametric option for curve reconstruction is feasible.

To fit functional linear models to our data, the R fda library[8] was used. As the package offers only limited functionality, the calculation of confidence intervals and the proportion of explained variance were done manually and can be found in the supplementary code.

**Model comparison**

To compare the standard linear models with functional ones, we selected the proportion of explained variance as a criterion. This information is included in the standard linear model output as multiple R-squared and can also be calculated for the functional linear model by dividing the sum of squared residuals (difference between the predicted value and actual value of the modeled variable to the power of two) by the sum of squared total variation (difference between the actual value of the modeled variable and the mean value of the modeled variable to the power of two) and subtracting the result from one. The resulting number takes values between 0 and 1 and expresses the proportion of the variance in the data explained by the model.

## Results

**Linear models – age at menarche**

To evaluate the novel analytical approach presented here, we first used standard epidemiological methods (linear regression models) to predict the age at menarche from data about height, weight, and mother's age at menarche of girls from the ELSPAC cohort. The best model, judged by the R-squared, created in this manner was able to explain 27% of the observed variance (Table 1).

**Table 1 Variance of age at menarche explained by the linear regression models.** *74 observations had to be excluded from these models because of missing maternal age at menarche

| Model | R-squared |
|---|---|
| Menarche ~ height at 11 y | 0.16 |
| Menarche ~ weight at 11 y | 0.21 |
| Menarche ~ BMI at 11 y | 0.12 |
| Menarche ~ height at 11 y + weight at 11 y | 0.22 |
| Menarche ~ height at 11 y + weight at 11 y + menarche mother* | 0.27 |
| Menarche ~ peak height velocity | 0.23 |
| Menarche ~ peak height velocity + menarche mother* | 0.22 |

## Functional linear models – age at menarche

The first functional linear model we constructed used growth curves estimated from longitudinal height measurements to predict the age at menarche of girls from the ELSPAC cohort (Figure 1). It showed that the girl's height from the interval approximately between 11 and 15 yr is negatively associated with age at menarche, whereas the interval from 16 to 18 yr shows a positive correlation between height and age at menarche. Furthermore, the shape of the coefficient suggests that height data starting approximately at 11 yr seem to be most important for predicting age at menarche. Furthermore, this single-variable model explained 34% of age at menarches` variance, which is notably higher than the best simple linear model.

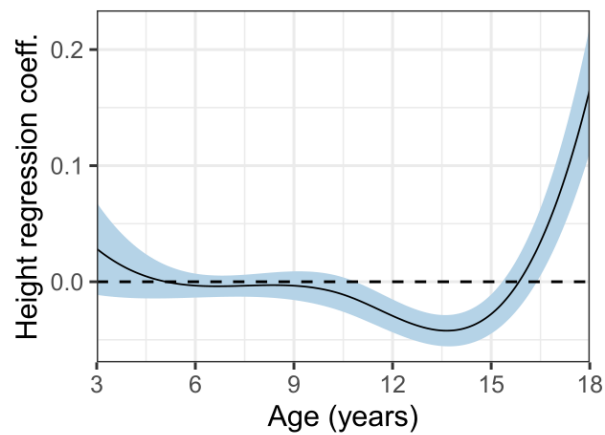

**Figure 1 Height-based functional linear model predicting ate at menarche.** The solid line represents a beta coefficient function and the light blue area marks the 95% confidence interval. The confidence interval areas that do not cross the horizontal line at zero can be considered statistically significant. The R-squared coefficient, representing the fraction of explained variance, is 0.34.

The second functional linear model was based on weight curves (Figure 2). It shows that the weight early in childhood, as well as in the period after 16 years of life, is positively associated with age at menarche. On the other hand, the period from 11 to 15 yr shows a negative association of weight with age at menarche. This model explains 28% of the observed variance, which makes it weaker than the height-based model, but still better than the best simple linear model.

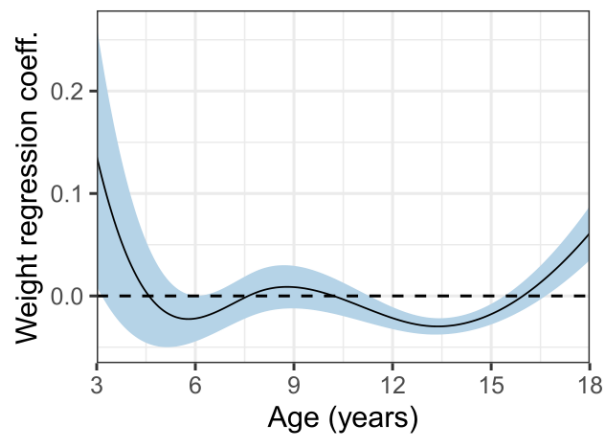

**Figure 2 Weight-based functional linear model predicting age at menarche.** The solid line represents a beta coefficient function, while the light blue area marks the 95% confidence interval. The confidence interval areas that do not cross the horizontal line at zero can be considered statistically significant. The R-squared coefficient, representing the fraction of explained variance, is 0.28.

The last functional linear model combined both growth and weight curves to predict age at menarche of girls from the ELSPAC cohort. While the coefficient curves are slightly different from the individual models, the conclusion remains similar. However, the R-squared of the combined model increased to 0.383, and thus, it outperforms the best linear regression model by 12% points in explaining the observed variance. In a next step, one can create a classical linear model using the functional model's information and include only height and weight at ages that seem most relevant. The resulting model (Age at Menarche ~ w(3 yr) + w(11 yr) + w(15 yr) + w(18 yr) + h(11 yr) + h(18 yr)) is similar to this one in terms of explained variance but still slightly worse (0.378).

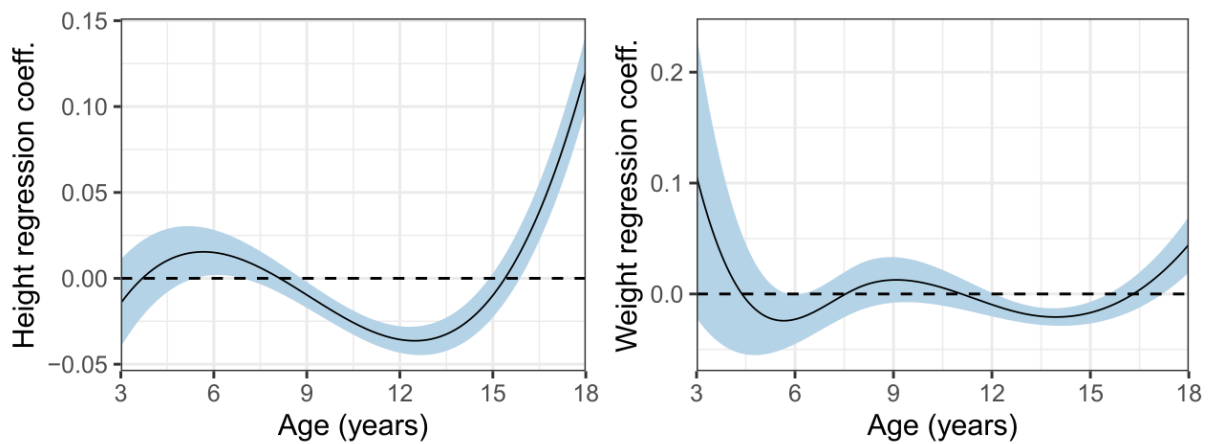

**Figure 3 Combined height/weight-based functional linear model predicting ate at menarche.** The solid line represents a beta coefficient function while the light blue area marks 95% confidence interval. The confidence interval areas that do not cross the horizontal line at zero can be considered statistically significant. The R-squared coefficient, representing the fraction of explained variance, is 0.383.

**Functional linear models – height at 18**

In addition to the suggestion that the development of a phenotype should be more informative of the development of another phenotype than just its currents state, the framework of the dynamic pathosome also suggests that a phenotype's past trajectory affects its future development. To test this hypothesis, we first used ELSPAC height data from 3, 5, 7, 11, and 15 years to construct functional linear models predicting height at 18 years (Figure 4). The sex of the child was also used as a covariate. The resulting model explains 89% of the variance. As shown in the previous section, a corresponding classic linear model with similar qualities can be created based on the same data using the functional model as a guide to select critical timepoints.

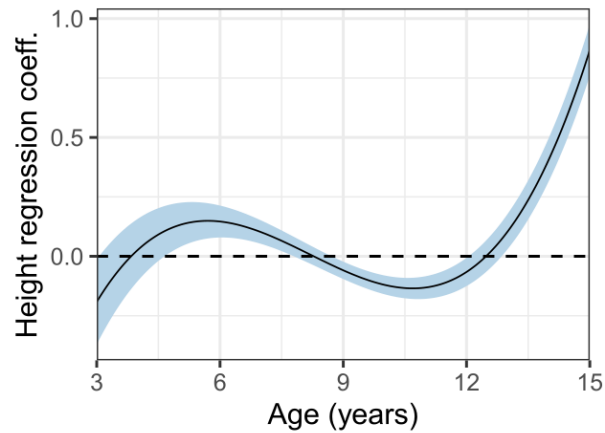

**Figure 4 Functional linear model predicting height at 18 years in ELSPAC cohort.** The solid line represents a beta coefficient function, while the dashed lines mark the 95% confidence interval. The confidence interval areas that do not cross the horizontal line at zero can be considered statistically significant. The R-squared coefficient, representing the fraction of explained variance, is 0.89. This model used sex as a covariate

To further test how the accuracy of the model is affected by a number of available time points, we utilized growth data from the BGS cohort. First, we construed functional models from sparse height data, i.e., measured only at 3, 5, 7, 11, and 15 years. This model explained 89% of the observed variance (Fig. 5a). In the next step, we constructed functional models from dense data, i.e., all heights measured in the BGS cohort until the subjects' 15 birthday. The model constructed in this fashion was stronger and explained 97% of the observed variance, demonstrating that more detailed phenotypic trajectories improve models' properties.

Furthermore, there is an apparent distinction between Figure 5a and Figure 5b. For the model with sparse data (Figure 5a), the coefficient curve seems flatter, and the confidence band is narrow. This is a non-intuitive but logical consequence of making data sparser by removing the data points. In general, fewer data means less variance, hence the slimmer confidence band. The removal of data points also causes flatness, as mentioned earlier. When using dense data, the model can find and emphasize the intervals that carry the majority of information used to predict the dependent variable. Thanks to the nature of continuous functional data such as height, the data points close to each other are also highly correlated; the correlation usually decreases with increasing distance. Therefore, if we remove a data point, a part of its information can still be found in the remaining data. The model can still extract the information but cannot pinpoint the associated time intervals used to predict the dependent variable as precisely as in the dense data model. One can view the coefficient curve obtained from sparse data as a "smoothed" version of the curve obtained from dense data.

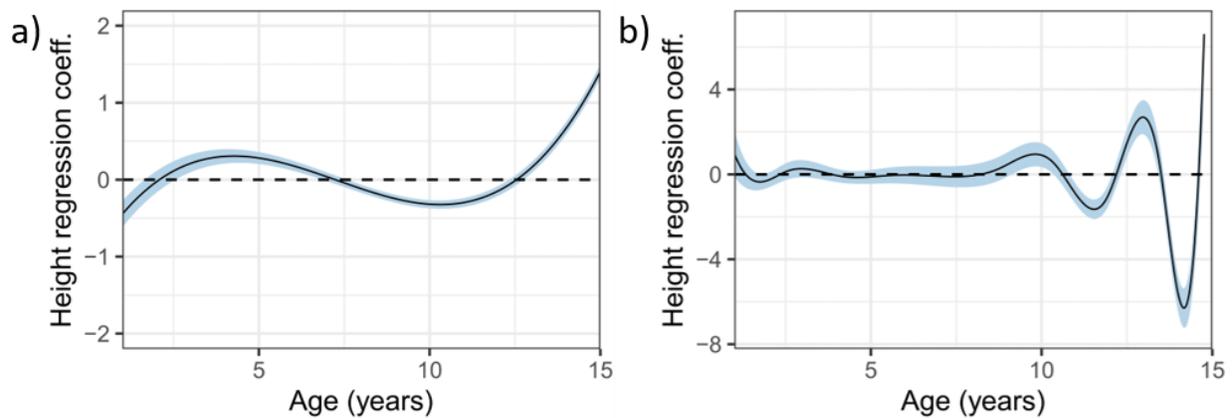

**Figure 5 Denser data improve the accuracy of functional models.** The solid line represents a beta coefficient function, while the dashed lines mark the 95% confidence interval. The confidence interval areas that do not cross the horizontal line at zero can be considered statistically significant. A) Functional model predicting height at 18 years in BGS cohort from height data measured at 3, 5, 7, 11, and 15 years. B) Functional model predicting height at 18 years in BGS cohort from height data measured at 25 time points from birth to 15. The R-squared coefficient, representing the fraction of explained variance, is 0.89 for the model in panel A and 0.97 for the model presented in panel B. These models did not use sex as a covariate.

One way to interpret the results presented in Fig 5b is that in the BGS cohort, at least 97% of the height variance at age 18 was already predetermined three years upfront, and the subsequent environmental stimuli accounted at best for 3% of the variance. To see how this trend developed in time, we constructed functional models from phenotypic trajectories ending at different ages and calculated their respective R-squared (Figure 6). First, we constructed functional models using only height data from the BGS cohort (Figure 6a). The variance explained by the model rose by almost 20% from the age of one to the age of three. Afterward, the explained variance stagnated around 50% until the age of 6. Then, in just two years, the explained variance improved by 10%. After a short slowing down, the models in the period of 9 to 15 years improved from explaining 60% of height's variance observed at age 18 to 97%.

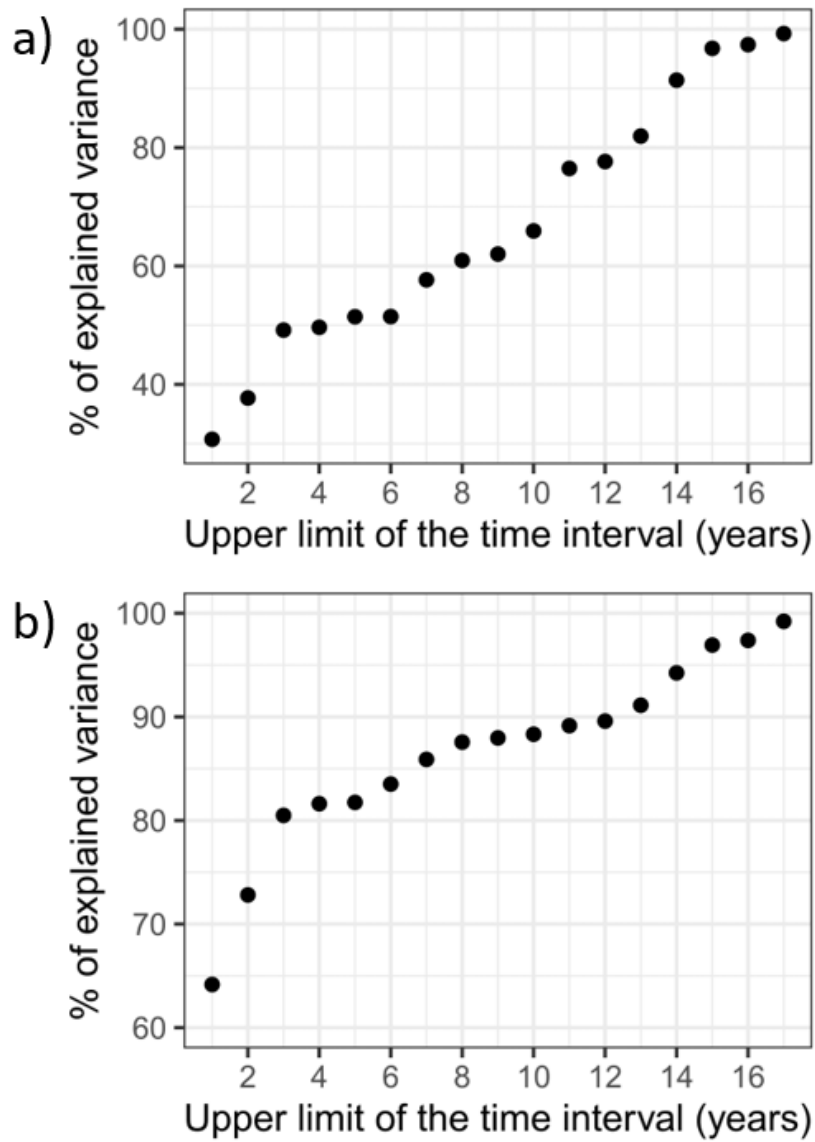

**Figure 6 Percentage of variance explained by phenotypic trajectory ending at different ages in BGS cohort.** Each dot on the graph represents the R-squared of a functional model predicting height at 18 from a growth curve starting at one year and ending at a specific age. A) Functional models based solely on height data. B) Functional models based on height data using sex as a covariate.

Second, we constructed functional models using sex as a covariate (Fig 6b). Results clearly showed that functional models with sex are much more precise at lower ages than models using just height data. However, the differences between models decreased rapidly with age, and at 15, the functional models with and without sex had approximately the same R-squared. Just as in the models without sex, variance explained by the models rose sharply at first, increasing by 15% in just two years. Surprisingly, the functional model based on a growth curve from 1-3 years already explained more than 80% of the observed variance of height at age 18. Properties of the models improved only slightly to the age of 5. Afterward, in the short period from 5-8 years, the variance explained by the models increased by more than 5%. From 8 to 12 years, the increase in the explained variance was much more gradual, accounting for approximately 2.5%. In the period from 12 to 15 years, the increase accelerated once again and the variance explained by the model improved by approximately 7.5%.

 As these models were adjusted by sex, they better reflect the part of the adult height's variance that is already "preprogrammed" at a certain age. While genetic and inherited factors are certainly responsible for a big part of this preprogramming, some of it is also caused by environmental factors that individuals at that age already encountered. One could even argue that a big part of the increase of models' precision with age is due to environmental effects which have occurred during developmental windows that have already closed. Therefore, intervals during which the variance explained by these models grows notably faster should reflect intervals where environmental inputs lose their ability to affect the final outcome. Such periods may point to the closing of distinct developmental windows allowing environmental influences to affect an individual's final height. Our results suggest the existence of three such developmental windows. The first of these windows starts to close one year or sooner and ends after reaching the third birthday. The second of these windows starts to close at five years and closes by eight, and the third one starts closing at 12 and closes by 15.

## Discussion

In this article, we have presented a novel pathosome-inspired approach based on functional linear models to analyze longitudinal epidemiological data. In essence, this approach differs from classic methods in that it analyzes entire phenotype trajectories, e.g., growth curves, instead of choosing one or a few specific timepoints. We have tested the pathosome-inspired approach on longitudinal data from the ELSPAC cohort and demonstrated that functional models outperform standard methods in explaining the observed variance in age at menarche. While our results also show that the functional model's information can be used to create classical linear models that are almost equally effective, creating such enhanced linear models still requires using functional models as a basis. Therefore functional linear models may serve as a valuable exploratory tool.

Furthermore, we have used the functional linear models to demonstrate that according to the prediction of the dynamic pathosome concept[1,2], the past phenotype trajectory predicts its future development. While the finding that growth patterns up to the 15th year predict the height at 18 years is hardly surprising, it is still remarkable that they explain at least 97% of the observed variance. Of course, it is still possible that denser datasets could make it possible to generate even more accurate models. In addition, we showed that even relatively short phenotypic trajectories from 1 to 3 years adjusted by sex explain more than 80% of the observed variance of height at age 18. We have also shown that the importance of sex for the model's accuracy decreases with the length of the phenotypic trajectory. This is likely because, over time, the difference in male and female growth patterns becomes more and more apparent from the growth curves themselves until all relevant information that could be gained by considering sex as a covariate is already contained in growth patterns. Besides, our results point to three periods during which the final height becomes quickly less malleable by the environmental influences possibly corresponding to the closure of specific developmental windows. Therefore, the approach presented here may constitute a novel method for the identification of candidate developmental windows.

Height and weight are well-known predictors of age at menarche[13,14], and thus, their analysis in this article served mainly as a demonstration of how functional linear models could be used in practice. Nevertheless, in addition to a simple replication of previous findings, our results have some practical implications. Many studies that are, for various reasons,  investigating age at menarche use BMI as a co-founder in their analysis[15–17]. While the choice of BMI instead of height or weight might seem straightforward, it may not always be the best option. Our results show that weight at age 11 was actually a better predictor of age at menarche than BMI in our cohort, see Table 1. However, a more important problem of such studies is that they use just a single number measured at a single timepoint. As we have shown, information about height, weight, or BMI from a single time point explains much less observed variance than their phenotypic trajectories or information about a few crucial timepoints selected by the help of functional linear models. Therefore, the practice of using information from a single timepoint unnecessarily decreases the precision of the final models. The fact that most of such studies used data from cohorts that measured height and weight repeatedly, but many authors deliberately chose not to use most of this information makes our finding particularly important as it highlights a relatively simple-to-implement solution that improves current practice in this field of research.

## References

1. Lenart, P., Scheringer, M. & Bienertová-Vašků, J. The Dynamic Pathosome: A Surrogate for Health and Disease. in *Explaining Health Across the Sciences* (eds. Sholl, J. & Rattan, S. I. S.) 271–288 (Springer International Publishing, 2020). doi:10.1007/978-3-030-52663-4_16.

2. Lenart, P., Scheringer, M. & Bienertova-Vasku, J. The Pathosome: A Dynamic Three-Dimensional View of Disease–Environment Interaction. *BioEssays* **41**, 1900014 (2019).

3. Baye, T. M., Abebe, T. & Wilke, R. A. Genotype–environment interactions and their translational implications. *Pers. Med.* **8**, 59–70 (2010).

4. Hunter, D. J. Gene–environment interactions in human diseases. *Nat. Rev. Genet.* **6**, 287–298 (2005).

5. European Longitudinal Study of Pregnancy and Childhood (ELSPAC). *Paediatr. Perinat. Epidemiol.* **3**, 460–469 (1989).

6. Piler, P. *et al.* Cohort Profile: The European Longitudinal Study of Pregnancy and Childhood (ELSPAC) in the Czech Republic. *Int. J. Epidemiol.* **46**, 1379–1379f (2017).

7. Ramsay, J. & Silverman, B. W. *Functional Data Analysis*. (Springer-Verlag, 1997). doi:10.1007/978-1-4757-7107-7.

8. Ramsay, J. O., Graves, S. & Hooker, G. *fda: Functional Data Analysis*. (2020).

9. Tuddenham, R. D. & Snyder, M. M. Physical growth of California boys and girls from birth to eighteen years. *Publ. Child Dev. Univ. Calif. Berkeley* **1**, 183–364 (1954).

10. R Core Team. *R: A Language and Environment for Statistical Computing*. (R Foundation for Statistical Computing, 2020).

11. Preece, M. A. & Baines, M. J. A new family of mathematical models describing the human growth curve. *Ann. Hum. Biol.* **5**, 1–24 (1978).

12. Elzhov, T. V., Mullen, K. M., Spiess, A.-N. & Bolker, B. *minpack.lm: R Interface to the Levenberg-Marquardt Nonlinear Least-Squares Algorithm Found in MINPACK, Plus Support for Bounds*. (2016).

13. Cooper, C., Kuh, D., Egger, P., Wadsworth, M. & Barker, D. Childhood growth and age at menarche. *Br. J. Obstet. Gynaecol.* **103**, 814–817 (1996).

14. Simmons, K. & Greulich, W. W. Menarcheal age and the height, weight, and skeletal age of girls age 7 to 17 years. *J. Pediatr.* **22**, 518–548 (1943).

15. Roberts, E., Fraser, A., Gunnell, D., Joinson, C. & Mars, B. Timing of menarche and self-harm in adolescence and adulthood: a population-based cohort study. *Psychol. Med.* **50**, 2010–2018.

16. Marks, K. J. *et al.* Exposure to phytoestrogens in utero and age at menarche in a contemporary British cohort. *Environ. Res.* **155**, 287–293 (2017).

17. Adair, L. S. Size at Birth Predicts Age at Menarche. *Pediatrics* **107**, e59–e59 (2001).

## Acknowledgments

We would like to thank Jan Koláček for his specialized course on functional data analysis that inspired the original idea that evolved into the approach presented in this article. The project was supported by the CETOCOEN PLUS (CZ.02.1.01/0.0/0.0/15_003/0000469) project of the Ministry of Education, Youth and Sports of the Czech Republic. The project was also supported by CETOCOEN EXCELLENCE Teaming 2 project supported by Horizon2020 (857560) and the Ministry of Education, Youth and Sports of the Czech Republic

(02.1.01/0.0/0.0/18_046/0015975). As well as by the RECETOX Research Infrastructure (LM2018121). D.K. was also supported by a specific research grant of Masaryk University (MUNI/A/1615/2020).

## Authors' contributions

P. L. formulated the research problem and interpreted the results. D. K. analyzed the data, J. B. V and M.S. supervised the project. L.K. managed the database with cohort data. All authors co-wrote the manuscript. P. L and D. K. contributed to the manuscript in equal measure.

## Competing interests

The authors declare no competing interests.

## Code availability

The example code to demonstrate use and calculation of functional linear models and R-squared in this study can be accessed at this [address](). Only BGS data is used in the code due to its unlimited availability.

## Data accessibility

Data from the ELPSAC used in this submission are available upon request to the project manager (elspac@recetox.muni.cz). However, restrictions apply to the availability of these data, which were used under license for the current study, and so are not publicly available. The process of requesting data is described at [http://www.elspac.cz/index-en.php?pg=professionals--partnership-establishment](). Data from the BGS cohort are freely available at [https://rdrr.io/cran/fda/man/growth.html]().